\documentclass[%
 aip,
 amsmath,amssymb,
 reprint,%
]{revtex4-1}

\usepackage{graphicx}
\usepackage{dcolumn}
\usepackage{bm}

\usepackage[utf8]{inputenc}
\usepackage[T1]{fontenc}
\usepackage{mathptmx}
\usepackage{etoolbox}

\makeatletter
\def\@email#1#2{%
 \endgroup
 \patchcmd{\titleblock@produce}
  {\frontmatter@RRAPformat}
  {\frontmatter@RRAPformat{\produce@RRAP{*#1\href{mailto:#2}{#2}}}\frontmatter@RRAPformat}
  {}{}
}%
\makeatother

\begin{document}

\preprint{AIP/123-QED}

\title[]{Correlation effects in Barkhausen noise and magnetic attenuation in soft ribbons}

\author{E. Puppin, M. Zani and E. Pinotti}

\affiliation{ 
Dipartimento di Fisica\\
Politecnico di Milano\\
P.zza Leonardo da Vinci 32\\
20133 Milano (ITALY)
}

\email{ezio.puppin@polimi.it}

\date{\today}

\begin{abstract}
The propagation of the effects connected with the occurrence of magnetization reversals in an amorphous ribbon of  Fe$_{63}$B$_{64}$Si$_{8}$Ni$_{15}$ has been investigated using a method based on two pickup coils separated by a distance variable between 2 an 40 mm. The ratio $R$ between the voltage signals induced in the coils contains information on the location where the magnetization reversal took place. This information can be extracted by knowing the function $V(x)$ which represents the attenuation of a signal generated by a reversal that took place at a distance x from the coil. A mathematical model for extracting this function starting from the histogram of the experimentally measured values of $R$ is presented. The attenuation function obtained in this way is relatively independent on the distance between the coils, and this is in strong support of the correctness of the model adopted. 
\end{abstract}

\maketitle

%

\section{Introduction}

The macroscopic properties of ferromagnetic materials are represented by the usual hysteresis loop which gives a thermodynamic description of the evolution of magnetization under the effect of an external field. On a microscopic level however the smooth and reproducible behavior of the hysteresis loop breaks down and strong discontinuities are observed. This jumpy nature of magnetism is well known since the original experiment of Barkhausen \cite{Bark} who observed that magnetization reversal takes place through a sequence of sharp jumps instead of being a continuous process as the smooth aspect of the hysteresis loop might suggest. Ever since Barkhausen noise (BN) has been widely investigated and in recent years the interest for this phenomenon increased, both for its relevance in statistical physics \cite{DZ} and for the applications in the characterization of magnetic metals in industrial applications \cite{Suvi}.

Actually, the complexity of the magnetization process is overwhelming and it is unlikely that an accurate modeling will be available soon. On the other hand, the available experimental techniques allow to collect large amounts of high quality data suitable for different kind of statistical analysis. One of the most prominent features of BN is its fractal nature. The probability distribution of the  jumps taking place during magnetization reversal has been largely investigated and there is a general consensus on the facts that this distribution can be described with a power law of this type: $P({\Delta}M) = {\Delta}M^{-\alpha}$. The term ${\alpha}$ is called critical exponent.

Another relevant issue related to the magnetization process is the propagation of magnetic effects through the material \cite{Pinotti}. Here we use the term “propagation” with the following meaning. Under the effect of the increasing external field magnetization jumps take place inside the material. During each jump a magnetic wall, which is initially pinned, is suddenly stripped off from its position and moves across the material up to its successive pinned configuration. In this way a certain portion of the sample undergoes a magnetization reversal. Even if this process involves only a small portion of the sample, its effects propagate through the material following the sudden variation of the magnetic flux associated with the magnetization reversal. Improving our understanding of the attenuation of these effects is a relevant issue in connection with the diagnostic applications of the Barkhausen noise.  

In a typical Barkhausen experiment a pick-up coil is placed close to the sample and the BN is measured as a series of well separated peaks. Each peak originates from a single magnetization jump which gives rise to an induced voltage signal in the pick-up coil. The amplitude of this peak depends both on the amplitude of the magnetization variation inside the sample (${\Delta}M$) and on the distance from the pick-up coil of the reversion region.

\section{Experimental apparatus}

\begin{figure}
\centering
\resizebox{0.45\textwidth}{!}{
  \includegraphics{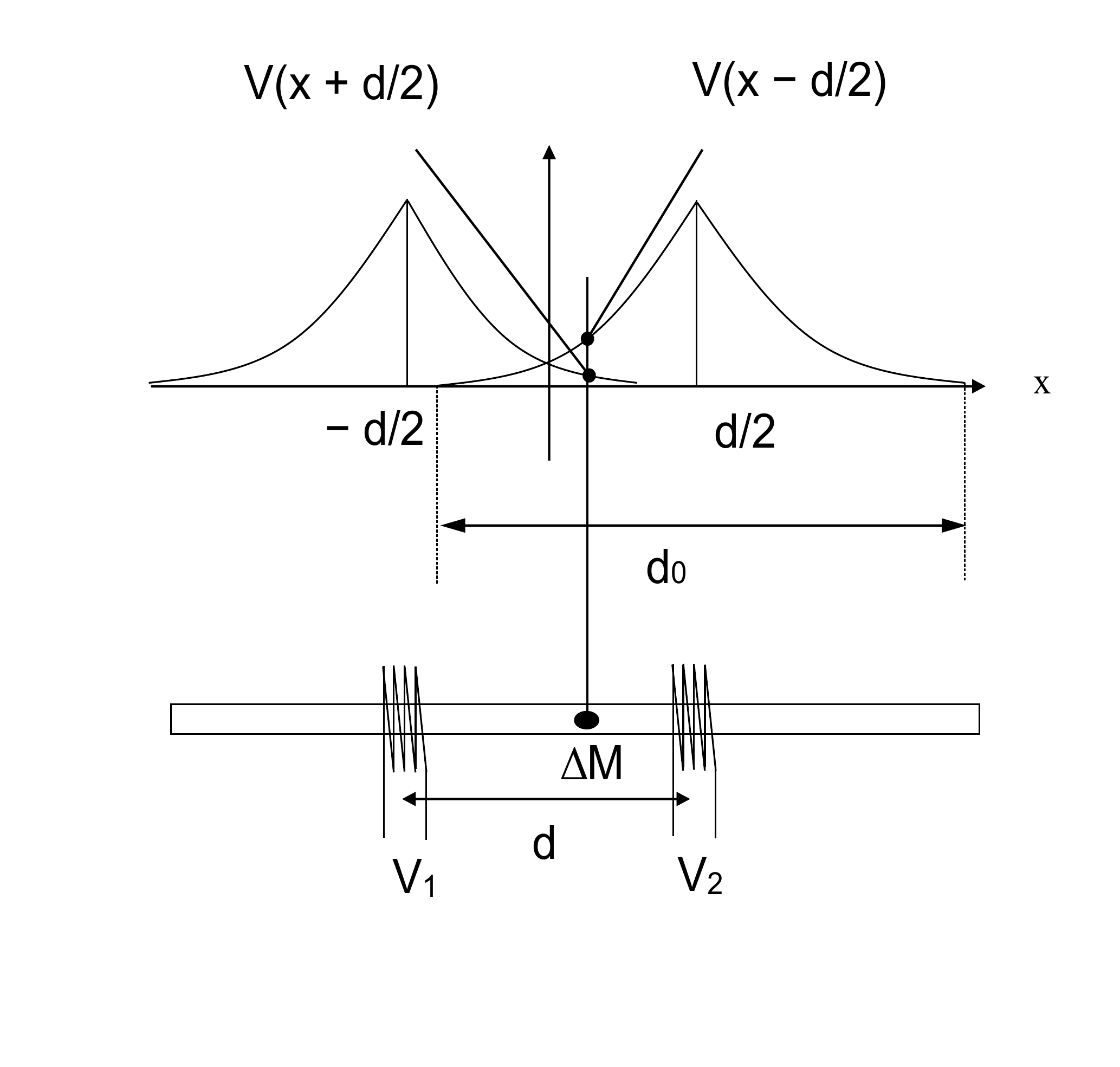}
}
\caption{\label{fig:F1} Experimental layout: 1) lower part: around the sample are placed two pickup coils and the magnetic field is applied along the ribbon direction; 2) upper part: the intensity of the voltage pulses depends both on the magnetization reversal amplitude ${\Delta}M$ and on the attenuation of the associated flux variation inside the material, represented by the unknown function $V(x)$.}
\end{figure}

The experimental apparatus is described in detail in Ref.~\onlinecite{RSIPuppin} and schematically represented in the lower part of Fig.~\ref{fig:F1}.  The sample used in this investigation is a ribbon of amorphous Fe$_{63}$B$_{64}$Si$_{8}$Ni$_{15}$ 100 mm long, 3 mm large and 20 ${\mu}m$ thick prepared by rapid quenching of the molten material. In order to ensure mechanical stability the ribbon has been glued on a stiff plastic support. This assembly has been placed within a magnetizing coil made of 1000 turns of copper capable to generate a magnetic field up to 1000 Oe. The two pickup coils used for detecting the induced BN can be moved along the sample length. Each coil is made of 1000 turns of a shielded copper wire (diameter 0.1 mm); the thickness of each coil is 0.5 mm. The whole apparatus is enclosed in a metallic box for shielding the electromagnetic noise. The coils can be moved one with respect to the other using a manipulator in order to change their relative distance without opening the shielding box. During our measurements the intermediate position of the two coils has been always kept in the middle of the ribbon. 

The acquisition is controlled by a PC which provides to the generation of the magnetic field and to the simultaneous detection of the two signals coming from the coils. In a typical acquisition the field performs a complete hysteresis cycle in 20 s. With an acquisition frequency of 50 kHz on each channel the number of sampling from each coil is 10$^{6}$ during one cycle. In this way it is possible to sample nearly 50000 peaks on each channel within one loop. All the data presented in the following come from a single hysteresis loop since the number of peaks is sufficient for the statistical accuracy needed.

Let us consider Fig.~\ref{fig:F2} where two small portions of the signals induced in the coils are shown. In the upper panel of Fig.~\ref{fig:F2} the distance $d$ between the coils was set at its minimum value ($d$ = 2 mm). Clearly the two signals are very similar and every single peak in one coil matches with an equivalent peak in the other coil. In the lower panel we see the case of $d$ = 20 mm. In this case the correlation is lower. In fact in the figure there are peaks detected only by one of the coils (such as peak A, C and F). For the peaks detected simultaneously by the two coils the amplitude is not the same due to the different attenuation.

For each peak generated within the coils it is possible to define the intensity which is simply the area below the peak. The area of an individual peak depends on two distinct factors: the true value of the magnetization reversal (${\Delta}M$) and the distance between the reversal location and the coil. We will assume that the size of the sample portion involved in these reversions is small with respect to the distance between the coils also in the smallest case of $d$ = 2 mm. This assumption is justified by the large number of peaks in one loop (25000 peaks in each branch of the loop) which corresponds to an average area of 10000 ${\mu}m^2$, to which corresponds a size in the order of 100 ${\mu}m^2$. Within this assumption it is possible to define the distance $x$ between the reversal place and the coil. 

In the present paper we will present a statistical analysis of a set of data taken with an experimental technique \cite{RSIPuppin} based on the idea of using two pick-up coils instead of one for measuring the BN. In this way it is possible to collect information on the attenuation inside the material of the effects produced by the magnetization reversal.

\section{Experimental results}

In order to understand the basic idea let us consider Fig.~\ref{fig:F1}. The coils are placed around the sample at a distance $d$. The $x$ coordinate travels along the sample length and the origin is placed midway between the coils. A magnetization reversal taking place in a location $x$ and having amplitude ${\Delta}M$ will generate two peaks having amplitude (defined as the area below the peak)  $V_{1}$ and $V_{2}$, both proportional to ${\Delta}M$. Furthermore, it is necessary to consider the attenuation of each signal. This attenuation is described by a function $V(x$) which will be a decreasing function of the distance between the reversal point and the coil. We have therefore: $V_{1}$ proportional to ${\Delta}M \times V(x + d/2)$ and $V_{2}$ to ${\Delta}M \times V(x - d/2)$. It can be assumed without loss of generality that the function $V(x)$ is normalized to 1, i.e., that $V(0) = 1$. It is also, for obvious symmetry reasons, $V(x) = V(-x)$. Let us define $R(x)$ as the ratio between the pulses generated in the two coils by the same magnetization reversal ${\Delta}M$:

\begin{figure}
\centering
\resizebox{0.35\textwidth}{!}{
  \includegraphics{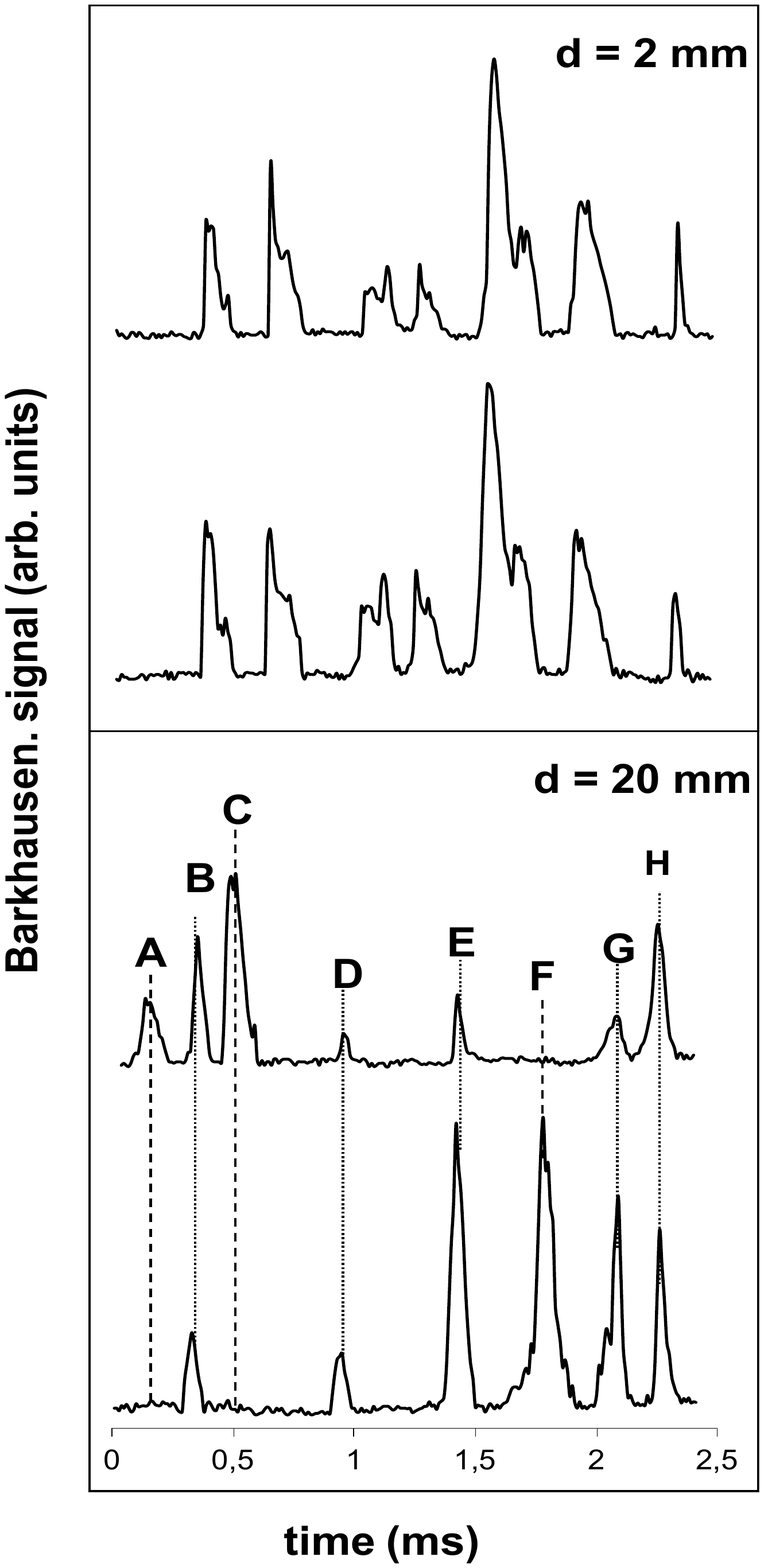}
}
\caption{\label{fig:F2} Voltage peaks from the two coils separated by a distance $d = 2$ mm in the upper panel and $d = 20$ mm in the lower panel. In the first case the two signal are practically equal whereas in the second case the correlation between the various peaks is smaller and the coincident peaks have significantly different amplitudes.}
\end{figure}

\begin{eqnarray}
R( x ) = \frac{{{\rm{\Delta }}M \times {V_1}( x )}}{{{\rm{\Delta }}M \times {V_2}( x )}} = \frac{{V( {x + \frac{d}{2}} )}}{{V( {x - \frac{d}{2}} )}}
\label{eq:E1}.
\end{eqnarray}

We see that $R$  does not depend on ${\Delta}M$. Given the experimental geometry $R(-x) = 1/R(x)$. In the following we will show how to extract information on $V(x)$ from the ensemble of the measured values of $R$.

The first step to perform is to evaluate the area of the various peaks in the two stream of data coming from the coils. This task is quite straightforward and it will not be discussed in detail. The second step is to identify the “corresponding” peaks in the two stream. Corresponding here means that we observe two peaks in the coils which are both originated from the same magnetization reversal. The criterion adopted is that two peaks are considered to be in coincidence if they fall within a certain time window. Also in this case the procedure can be simply accomplished by the analysis program and its details will be not discussed. The only relevant point to note is that the results we obtain do not critically depends on the details of these two operations and that the adopted algorithms are quite robust.

\begin{figure}
\centering
\resizebox{0.40\textwidth}{!}{
  \includegraphics{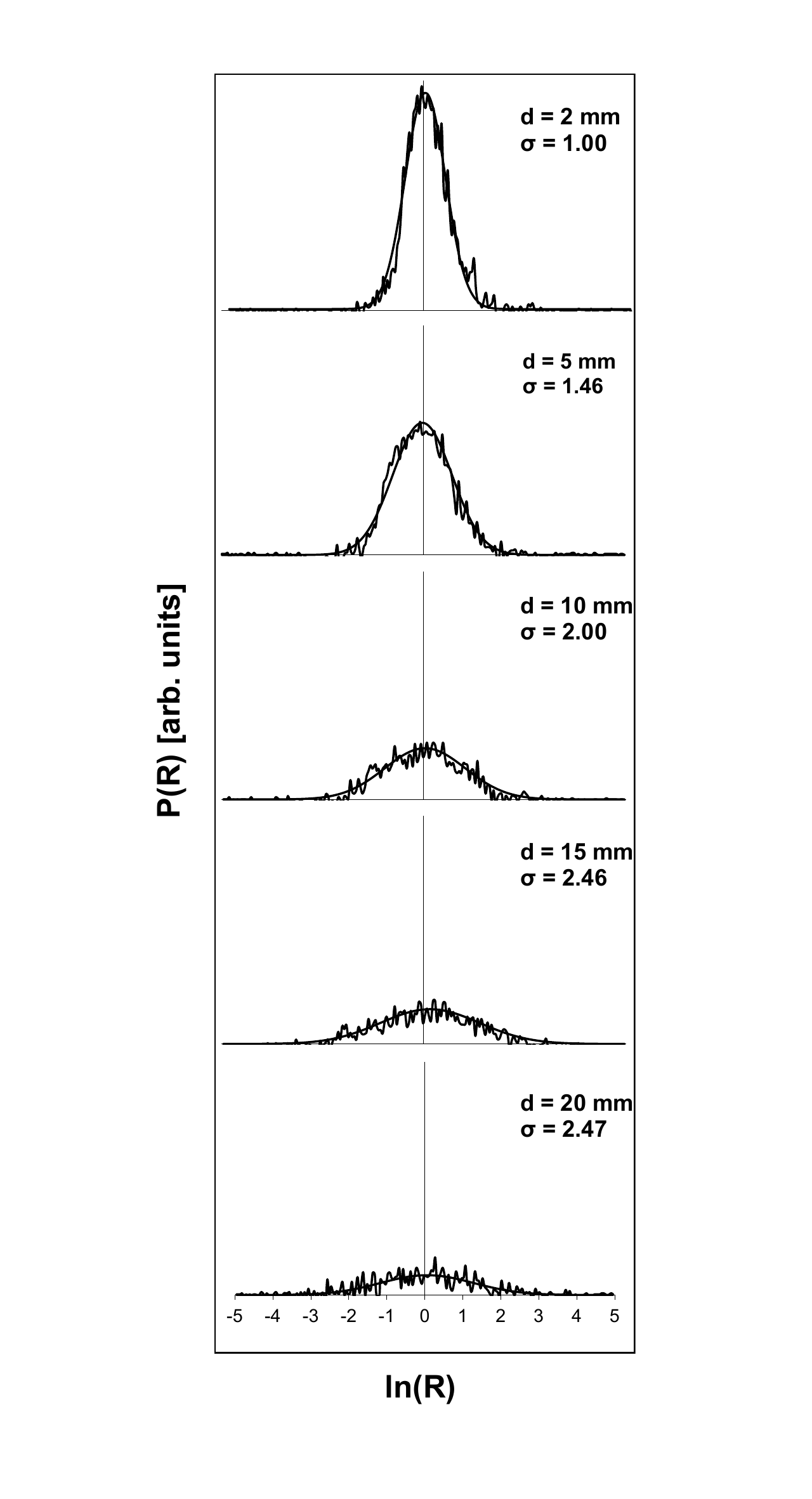}
}
\caption{\label{fig:F3} Experimental values of $R$, the ratio of the peaks detected by the two coils placed at increasing distances. The solid smooth line is a Gaussian fit of the experimental data.}
\end{figure}

The output of this analysis is an ensemble of values of $R$, the peak area ratio. Since the number of coincident peaks is sufficiently high it is possible to plot the histogram of these number which, by definition, is probability distribution $P(R)$ of obtaining a particular value of $R$. A series of these distributions, measured for different values of the distance between the coils), is shown in Fig.~\ref{fig:F3}. In the horizontal axis it is shown the logarithm of $R$. The first point to consider is the total number of points of each histogram. When the coils are very close each other the number of points is high. Actually, in each loop the number of peaks is much higher, in the order of 50000. However, in order to avoid artifacts we set a high thresholds for accepting the peaks and for deciding if they are coincident or not. This selection lower the number of events to 6573 for the case of $d = 2$ mm. The same selection criteria have been adopted also in the other cases shown in Fig.~\ref{fig:F3}. 

Above a certain separation between the coils the statistical distribution $P(R)$ does not depend on the distance itself. This indicates that the coincident peaks observed are not due to the same magnetization reversal but, instead, are produced by separate events. The distance above which this occur is in the order of 40 mm. For this reason we decided to consider the distribution at $d = 40$ mm as representative of the background of those events originated by uncorrelated magnetization reversion taking placed simultaneously. This background is present in all distributions and therefore we subtracted the distribution for $d = 40$ mm from all the others. The distributions shown in Fig.~\ref{fig:F3} have been already corrected by removing this background.

In Fig.~\ref{fig:F3} the corrected probability distributions of $R$ is shown in a semi-log plot and in this representation they appear to be symmetrical. This symmetry reflects the symmetry of the experiment as illustrated in Fig.~\ref{fig:F1}. A reversal taking place at a distance $x$ from the origin has the same probability to occur of the same reversal at a distance $-x$. On the other hand the ratio $R$ in position $x$ will become $1/R$ in position $-x$. A representation in a semi-log plot assures that $R$ and $1/R$ have the same probability to occur. 

Another interesting observation is that the probability distribution of $R$ is nearly Gaussian in a semi-log plot. The width of the distribution increases for increasing $d$ and the corresponding values of the FWHM are indicated with $w$ in each panel.

\section{The attenuation function}

In the following it will be discussed how to recover the attenuation function $V(x)$ of the voltage pulses detected by the coils starting from the experimentally determined probability distribution $P(R)$ of their ratio $R$ whose meaning is the following: $P(R)dR$ represent the number of reversal giving rise to a value of $R$ within an interval $dR$. As seen before in Eq.~(\ref{eq:E1}) the events giving rise to a ratio $R$ occur in a well defined position $x$. For this reason the number of events within $dR$ is equal to the number of magnetization reversals taking place within the corresponding portion $dx$ of the sample. If we define $p(x)$ as the linear density of reversals taking place inside the sample, we can conclude that: 

\begin{eqnarray}
P(R)dR = p(x)dx
\label{eq:E2}
\end{eqnarray}

Since, as we have seen, the distributions are symmetrical in a semi-log plot, it is advisable to operate with the logarithms of the two functions $V(x)$ and $R(x)$:

\begin{eqnarray}
w(x) = ln[{V(x)}]
\label{eq:E3}
\end{eqnarray}

\begin{eqnarray}
z(x) = ln[{R(x)}] = ln \left[{V\left({x + \frac{d}{2}}\right)}\right] - ln\left[{V\left({x - \frac{d}{2}}\right)}\right] = \nonumber\\
w\left({x + \frac{d}{2}}\right) - w\left({x - \frac{d}{2}}\right)
\label{eq:E4}
\end{eqnarray}

The function $w(x)$ can be expanded in power series:

\begin{eqnarray}
w(x) = \mathop \sum \limits_{n = 0}^\infty  {a_n}{x^n}
\label{eq:E5}
\end{eqnarray}

According to the definition of Eq.~(\ref{eq:E3}) and the condition $V(0) = 1$ it must be $w(0) = 0$ and therefore $a_0 = 0$. Furthermore, the function $w(x$) has the same symmetry as $V(x$), that is, $w(-x) = w(x)$. Therefore, only the terms with an even index are present in the summation:

\begin{eqnarray}
w(x) = \mathop \sum \limits_{k = 1}^\infty  {a_{2k}}{x^{2k}} = {a_2}{x^2} + {a_4}{x^4} + {a_6}{x^6} +  \ldots 
\label{eq:E6}
\end{eqnarray}

It is now possible to expand also the function $z(x)$ by replacing the variables $x+d/2$ and $x-d/2$ in Eq.~(\ref{eq:E5}):

\begin{eqnarray}
z( x ) = w\left({x + \frac{d}{2}} \right) - w\left( {x - \frac{d}{2}} \right) = \nonumber\\
\mathop \sum \limits_{k = 1}^\infty  {a_{2k}}{\left( {x + \frac{d}{2}} \right)^{2k}} - \mathop \sum \limits_{k = 1}^\infty  {a_{2k}}{\left( {x - \frac{d}{2}} \right)^{2k}} = \nonumber\\
\mathop \sum \limits_{k = 1}^\infty  {a_{2k}}\left[ {{{\left( {x + \frac{d}{2}} \right)}^{2k}} - {{\left( {x - \frac{d}{2}} \right)}^{2k}}} \right]
\label{eq:E7}
\end{eqnarray}

It is easy to verify that while the function $w(x$) contains only the even powers and therefore is even, the $z(x$) resulting from expression (\ref{eq:E6}) contains only the odd powers of $x$ and therefore is odd: $z(-x) = -z(x)$, as required from its definition (\ref{eq:E3}) and given that $R(-x) = 1/R(x)$.

From (\ref{eq:E6}) the function $z(x)$ is expressed by a series of powers that will be completely identified once the coefficients $a_{2k}$ are obtained. In order to find them, one must resort to the experimental data which give the probability density $p(z)$ of the variable $z$ and assuming the probability density $p(x)$ that a jump occurs at the $x$ coordinate. The probability of obtaining a certain $z$ is related to the probability that the jump occurs at the $x$ coordinate by the relationship:

\begin{eqnarray}
p( z )dz = p( x )dx
\label{eq:E8}
\end{eqnarray}

The two terms of this equation are integrated into their respective domains. Given the symmetry of the $z$ function, it is sufficient to integrate only on the right half of the $x$ axis:

\begin{eqnarray}
\mathop \int \nolimits_0^z p( z )dz =  - \mathop \int \nolimits_0^x p( x )dx
\label{eq:E9}
\end{eqnarray}

The minus appears because after Eq.~(\ref{eq:E1}) in the region $x	\geqslant0$ the amplitude of $z$ decreases as $x$ increases. Once the functions $p(z$) and $p(x$) are known, these integrals provide the function $z(x)$ which will in turn be expanded in series:

\begin{eqnarray}
z( x ) = \mathop \sum \limits_{n = 1}^\infty  {C_n}{x^n}
\label{eq:E10}
\end{eqnarray}

This time the coefficients $C_n$ are known, because the function $z(x)$ obtained from the integral of Eq.~(\ref{eq:E9}) is known. The series of Eq.~(\ref{eq:E10}) is compared term by term with the series of Eq.~(\ref{eq:E7}), so that from the known coefficients $C_n$ we can derive the unknown coefficients $a_{2k}$. From the latter we build the function $w(x$) through the series of Eq.~(\ref{eq:E5}) and finally the desired function $V(x) = exp(w(x))$. If the convergence of the series is rapid, it will be sufficient to consider a limited number of terms and the calculation can be done analytically. However, nothing stands in the way of implementing a numerical calculation and therefore considering any number of terms.
In the experimental results described in this paper, the probability density is Gaussian: normalizing to the total number of jumps we have: 

\begin{eqnarray}
p( z ) = \frac{1}{{\sqrt {2\pi } \sigma }}exp\left({ - \frac{{{z^2}}}{{2{\sigma ^2}}}}\right)
\label{eq:E11}
\end{eqnarray}

As for the probability density $p(x)$, it can be assumed to be uniform. The part of the strip that is measured by the spaced coils has a length equal to the sum $d+d_0$ where $d$ is the distance between the coils and $d_0$ is the extension of the region around each coil such that the events taking place inside it are detected by the coil (see Fig.~\ref{fig:F1} for visualising its meaning). In the following calculations we assumed $d_0$ = 40 mm but it must be emphasized that, as far as the determination of the attenuation function $V(x)$ is concerned, the exact value assigned to this parameter is non particularly relevant. Therefore:

\begin{eqnarray}
p( x ) = \frac{1}{{d + {d_0}}}
\label{eq:E12}
\end{eqnarray}

Eq.~(\ref{eq:E8}) thus becomes:

\begin{eqnarray}
\mathop \int \nolimits_0^z \frac{1}{{\sqrt {2\pi }\sigma }}exp\left({ - \frac{{{z^2}}}{{2{\sigma ^2}}}}\right)dz =  - \mathop \int \nolimits_0^x \frac{1}{{d + {d_0}}}dx
\label{eq:E13}
\end{eqnarray}

By changing variable variable $z^2 = 2{\sigma}t^2$ it transforms as:

\begin{eqnarray}
\frac{2}{{\sqrt \pi  }}\mathop \int \nolimits_0^{z/\sqrt 2 \sigma } {e^{ - {t^2}}}dt =  - \frac{{2x}}{{d + {d_0}}}
\label{eq:E14}
\end{eqnarray}

\begin{figure}
\centering
\resizebox{0.43\textwidth}{!}{
  \includegraphics{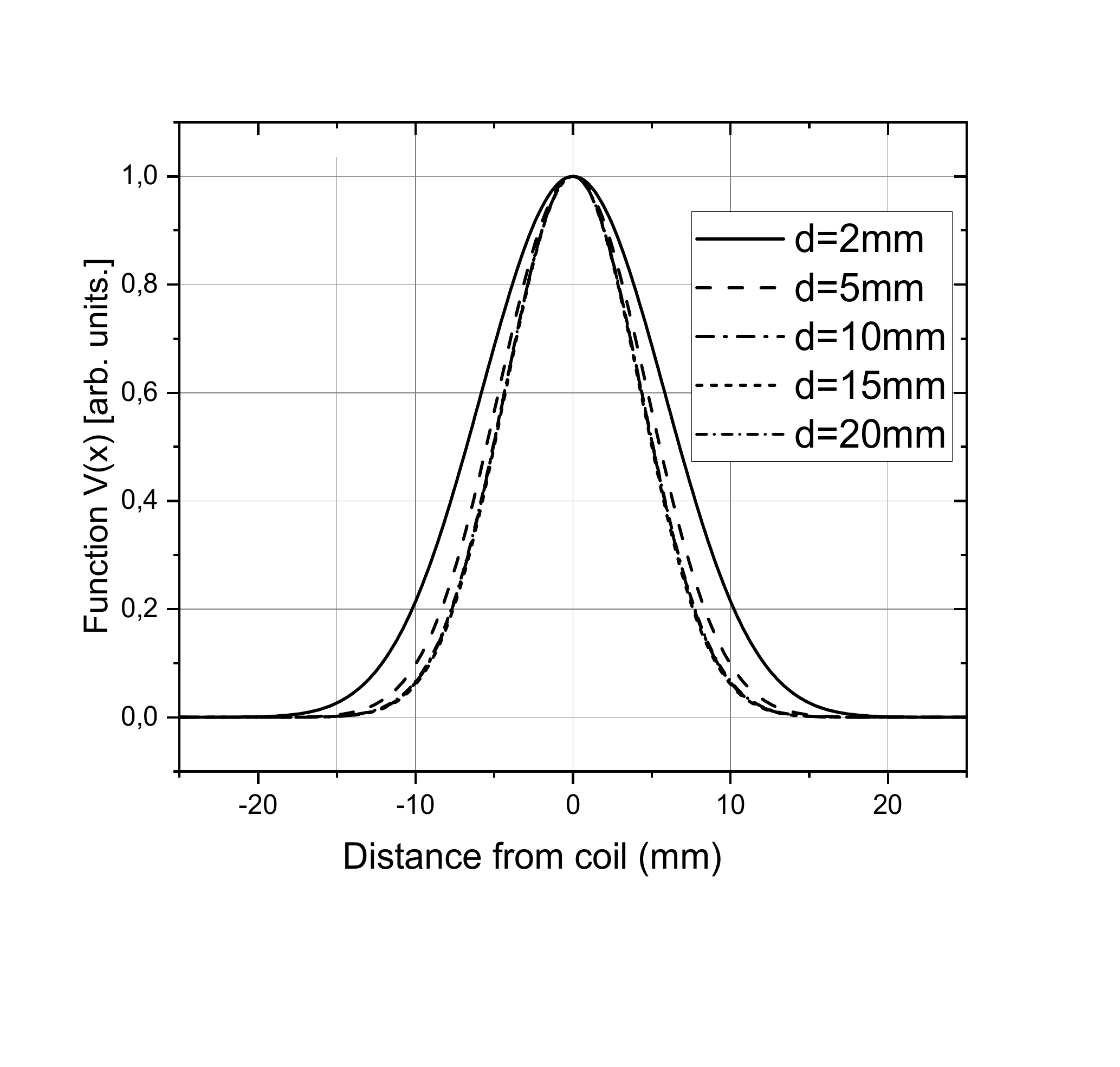}
}
\caption{\label{fig:F4} Attenuation function V(x) calculated for different values of the distance $d$ between the coils. }
\end{figure}

The expression on the left is nothing but the error function:

\begin{eqnarray}
{erf}\left({\frac{z}{{2\sigma }}}\right) =  - \frac{{2x}}{{d + {d_0}}}
\label{eq:E15}
\end{eqnarray}

And therefore:

\begin{eqnarray}
z = -2{\sigma}{erf}^{-1}\left(\frac{{2x}}{{d + {d_0}}}\right)
\label{eq:E16}
\end{eqnarray}

here with $erf^{-1}$  we mean the inverse function of the error function. The series expansion of this function is known \cite{Strecock}:

\begin{eqnarray}
{erf}^{ - 1}( y ) = \mathop \sum \limits_{n = 1}^\infty  {C_n}{y^{2n - 1}} = {C_1}y + {C_2}{y^3} + {C_3}{y^5} + {C_4}{y^7} +  \ldots 
\label{eq:E17}
\end{eqnarray}

The first 200 $C_n$ coefficients are tabulated in Ref.~\onlinecite{Strecock} and the first four orders are reported in Table.~\ref{tab:table1}. 

\begin{table}
\caption{\label{tab:table1}Coefficients of the series expansion of $erf^{-1}(y)$}
\begin{ruledtabular}
\begin{tabular}{lr}
Order&Coefficient\\
\hline
1 & 0.8862\\
2 & 0.2320\\
3 & 0.1276\\
4 & 0.0866\\
\end{tabular}
\end{ruledtabular}
\end{table}

Therefore the series expansion of the function $z(x) = ln(R(x))$ obtained from its Gaussian probability density and a uniform probability density for $x$ is:

\begin{eqnarray}
z( x ) = -2{\sigma}\mathop \sum \limits_{n = 1}^\infty  {C_n}\left({\frac{{2x}}{{d + {d_0}}}}\right)^{2n-1}
\label{eq:E18}
\end{eqnarray}

The previous expression can now be numerically evaluated since all its terms are known: the $C_n$ coefficients, the experimental parameters $d$ and $d_0$, the experimentally determined standard deviation of the various distributions shown in Fig.~\ref{fig:F3}. By comparing term by term this series with Eq.~(\ref{eq:E7}) we obtain the $a_2k$ as a function of $d$ and $d_0$ and then we can calculate the series expression of $w(x)$ (see Eq.~(\ref{eq:E6})) and finally $V(x$) (see Eq.~(\ref{eq:E3})). The calculation is not reported here for simplicity. Only the final result is reported in graphical form in Fig.~\ref{fig:F4}, i.e. the form of the function $V(x)$ obtained from the experimental data statistics at the various coil distances. As can be seen, despite the distance between the coils varies from 2 to 20 and the distribution variance changes from 1.08 to 2.56, the shape of the $V(x)$ function remains almost unchanged, as expected.

\section{Summary}

In the present paper we presented a model for interpreting Barkhausen effect data acquired using two coils. To each magnetization reversal taking place inside the sample the relevant information is represented by the ratio $R$ of the voltage peaks induced in the coils. A mathematical model for deriving the attenuation function $V(x)$ for these signals has been presented. 
By applying the procedure to different set of data, corresponding to very different values of the distance between the coils (from 2 to 40 mm) we derive the same attenuation function, with small variations for the different distances. This indicates that the procedure adopted is correct.
It is interesting to observe that the attenuation function is a polynomial containing four terms, with even power. We also performed numerical calculations (non reported here for space reasons) by trying to fit the experimental data with other type of functional dependence of the attenuation function (exponential, Gaussian and polynomial) and also in this case the best fit was obtained with the polynomial. 
Due to the complexity of the problem it is hard to give an explanation for this dependency of the signal attenuation on distance.

\section{Data Availability Statement}

The data that support the findings of this study are available from the corresponding author upon reasonable request.

\nocite{*}
\bibliography{Correlation}

\providecommand{\noopsort}[1]{}\providecommand{\singleletter}[1]{#1}%
\begin{thebibliography}{6}%
\makeatletter
\providecommand \@ifxundefined [1]{%
 \@ifx{#1\undefined}
}%
\providecommand \@ifnum [1]{%
 \ifnum #1\expandafter \@firstoftwo
 \else \expandafter \@secondoftwo
 \fi
}%
\providecommand \@ifx [1]{%
 \ifx #1\expandafter \@firstoftwo
 \else \expandafter \@secondoftwo
 \fi
}%
\providecommand \natexlab [1]{#1}%
\providecommand \enquote  [1]{``#1''}%
\providecommand \bibnamefont  [1]{#1}%
\providecommand \bibfnamefont [1]{#1}%
\providecommand \citenamefont [1]{#1}%
\providecommand \href@noop [0]{\@secondoftwo}%
\providecommand \href [0]{\begingroup \@sanitize@url \@href}%
\providecommand \@href[1]{\@@startlink{#1}\@@href}%
\providecommand \@@href[1]{\endgroup#1\@@endlink}%
\providecommand \@sanitize@url [0]{\catcode `\\12\catcode `\$12\catcode
  `\&12\catcode `\#12\catcode `\^12\catcode `\_12\catcode `\%12\relax}%
\providecommand \@@startlink[1]{}%
\providecommand \@@endlink[0]{}%
\providecommand \url  [0]{\begingroup\@sanitize@url \@url }%
\providecommand \@url [1]{\endgroup\@href {#1}{\urlprefix }}%
\providecommand \urlprefix  [0]{URL }%
\providecommand \Eprint [0]{\href }%
\providecommand \doibase [0]{http://dx.doi.org/}%
\providecommand \selectlanguage [0]{\@gobble}%
\providecommand \bibinfo  [0]{\@secondoftwo}%
\providecommand \bibfield  [0]{\@secondoftwo}%
\providecommand \translation [1]{[#1]}%
\providecommand \BibitemOpen [0]{}%
\providecommand \bibitemStop [0]{}%
\providecommand \bibitemNoStop [0]{.\EOS\space}%
\providecommand \EOS [0]{\spacefactor3000\relax}%
\providecommand \BibitemShut  [1]{\csname bibitem#1\endcsname}%
\let\auto@bib@innerbib\@empty
\bibitem [{\citenamefont {Barkausen}(1919)}]{Bark}%
  \BibitemOpen
  \bibfield  {author} {\bibinfo {author} {\bibfnamefont {H.}~\bibnamefont
  {Barkausen}},\ }\href@noop {} {\bibfield  {journal} {\bibinfo  {journal} {Z.
  Phys.}\ }\textbf {\bibinfo {volume} {20}},\ \bibinfo {pages} {401} (\bibinfo
  {year} {1919})}\BibitemShut {NoStop}%
\bibitem [{\citenamefont {Durin}\ and\ \citenamefont {Zapperi}(2005)}]{DZ}%
  \BibitemOpen
  \bibfield  {author} {\bibinfo {author} {\bibfnamefont {G.}~\bibnamefont
  {Durin}}\ and\ \bibinfo {author} {\bibfnamefont {S.}~\bibnamefont
  {Zapperi}},\ }\bibfield  {title} {\enquote {\bibinfo {title} {The barkhausen
  effects},}\ }in\ \href@noop {} {\emph {\bibinfo {booktitle} {The Science of
  Hysteresis}}},\ \bibinfo {editor} {edited by\ \bibinfo {editor}
  {\bibfnamefont {G.}~\bibnamefont {Bertotti}}\ and\ \bibinfo {editor}
  {\bibfnamefont {I.}~\bibnamefont {Mayergoyz}}}\ (\bibinfo  {publisher}
  {Elsevier},\ \bibinfo {address} {Amsterdam},\ \bibinfo {year}
  {2005})\BibitemShut {NoStop}%
\bibitem [{\citenamefont {Santa‑aho}\ \emph {et~al.}(2019)\citenamefont
  {Santa‑aho}, \citenamefont {Laitinen}, \citenamefont {Sorsa},\ and\
  \citenamefont {Vippola}}]{Suvi}%
  \BibitemOpen
  \bibfield  {author} {\bibinfo {author} {\bibfnamefont {S.}~\bibnamefont
  {Santa‑aho}}, \bibinfo {author} {\bibfnamefont {A.}~\bibnamefont
  {Laitinen}}, \bibinfo {author} {\bibfnamefont {A.}~\bibnamefont {Sorsa}}, \
  and\ \bibinfo {author} {\bibfnamefont {M.}~\bibnamefont {Vippola}},\
  }\href@noop {} {\bibfield  {journal} {\bibinfo  {journal} {J. of
  Nondestructive Evaluation}\ }\textbf {\bibinfo {volume} {38}},\ \bibinfo
  {pages} {94} (\bibinfo {year} {2019})}\BibitemShut {NoStop}%
\bibitem [{\citenamefont {Pinotti}\ and\ \citenamefont
  {Puppin}(2011)}]{Pinotti}%
  \BibitemOpen
  \bibfield  {author} {\bibinfo {author} {\bibfnamefont {E.}~\bibnamefont
  {Pinotti}}\ and\ \bibinfo {author} {\bibfnamefont {E.}~\bibnamefont
  {Puppin}},\ }\href@noop {} {\bibfield  {journal} {\bibinfo  {journal} {J.
  Appl. Phys.}\ }\textbf {\bibinfo {volume} {109}},\ \bibinfo {pages} {073909}
  (\bibinfo {year} {2011})}\BibitemShut {NoStop}%
\bibitem [{\citenamefont {Puppin}\ \emph {et~al.}(2001)\citenamefont {Puppin},
  \citenamefont {Zani}, \citenamefont {Vallaro},\ and\ \citenamefont
  {Venturi}}]{RSIPuppin}%
  \BibitemOpen
  \bibfield  {author} {\bibinfo {author} {\bibfnamefont {E.}~\bibnamefont
  {Puppin}}, \bibinfo {author} {\bibfnamefont {M.}~\bibnamefont {Zani}},
  \bibinfo {author} {\bibfnamefont {D.}~\bibnamefont {Vallaro}}, \ and\
  \bibinfo {author} {\bibfnamefont {A.}~\bibnamefont {Venturi}},\ }\href@noop
  {} {\bibfield  {journal} {\bibinfo  {journal} {Rev. Sci. Instr.}\ }\textbf
  {\bibinfo {volume} {72}},\ \bibinfo {pages} {2058} (\bibinfo {year}
  {2001})}\BibitemShut {NoStop}%
\bibitem [{\citenamefont {Strecock}(1968)}]{Strecock}%
  \BibitemOpen
  \bibfield  {author} {\bibinfo {author} {\bibfnamefont {A.~J.}\ \bibnamefont
  {Strecock}},\ }\href@noop {} {\bibfield  {journal} {\bibinfo  {journal}
  {Math. Comp.}\ }\textbf {\bibinfo {volume} {22}},\ \bibinfo {pages} {44}
  (\bibinfo {year} {1968})}\BibitemShut {NoStop}%
\end{thebibliography}%

\end{document}